\documentclass[aps,prb,amssymb,showpacs,twocolumn,floatfix]{revtex4}
\usepackage{graphicx}
\usepackage{bm}

\begin{document}

\title{Evolution of the structure of amorphous ice - from low-density
  amorphous (LDA) through high-density amorphous (HDA) to very high-density
  amorphous (VHDA) ice. }
\author{R.~Marto\v{n}\'ak} 
\altaffiliation[Permanent address: ]{Department of Physics (FEI), 
  Slovak University of Technology, Ilkovi\v{c}ova 3, 812 19 Bratislava,
  Slovakia} 
\email{martonak@phys.chem.ethz.ch}
\affiliation{Computational Science, Department of Chemistry and Applied
  Biosciences, ETH Zurich, USI Campus, Via Giuseppe Buffi 13, CH-6900
  Lugano, Switzerland} 
\author{D.~Donadio}
\affiliation{Computational Science, Department of Chemistry and Applied
  Biosciences, ETH Zurich, USI Campus, Via Giuseppe Buffi 13, CH-6900
  Lugano, Switzerland} 
\author{M.~Parrinello}
\affiliation{Computational Science, Department of Chemistry and Applied
  Biosciences, ETH Zurich, USI Campus, Via Giuseppe Buffi 13, CH-6900
  Lugano, Switzerland} 
\date{\today}

\begin{abstract}
  We report results of molecular dynamics simulations of amorphous ice for
  pressures up to 22.5 kbar. The high-density amorphous ice (HDA) as
  prepared by pressure-induced amorphization of I$_{\rm h}$ ice at $T=80$ K
  is annealed to $T=170$ K at various pressures to allow for relaxation.
  Upon increase of pressure, relaxed amorphous ice undergoes a pronounced
  change of structure, ranging from the low-density amorphous ice (LDA) at
  $p=0$, through a continuum of HDA states to the limiting very
  high-density amorphous ice (VHDA) regime above 10 kbar. The main part of
  the overall structural change takes place within the HDA megabasin, which
  includes a variety of structures with quite different local and
  medium-range order as well as network topology and spans a broad range of
  densities. The VHDA represents the limit to densification by adapting the
  hydrogen-bonded network topology, without creating interpenetrating
  networks. The connection between structure and metastability of various
  forms upon decompression and heating is studied and discussed. We also
  discuss the analogy with amorphous and crystalline silica. Finally, some
  conclusions concerning the relation between amorphous ice and supercooled
  water are drawn.
\end{abstract}

\pacs{64.70.Kb, 61.43.Er, 02.70.Ns, 07.05.Tp} 
\maketitle 

\section{Introduction}
\label{sec:introduction}

Properties of amorphous solid water at low temperatures continue to attract
the attention of both theory and experiment.  It has been known for a long
time that at least two distinct amorphous forms of ice exist. High-density
amorphous ice (HDA) is prepared by compression of ordinary I$_{\rm h}$ ice
to 12 kbar \cite{mishima1} and when recovered at ambient pressure it has a
density of $\sim$ 1.17 g/cm$^3$. Upon isobaric heating to 117 K, the
density drops considerably and a second form is found, called low-density
amorphous ice (LDA)\cite{mishima1} with a density $\sim$ 0.94 g/cm$^3$. The
transition between LDA and HDA can also be induced by pressure
\cite{mishima2,mishima3}. Interest in this system is enhanced by the
possible existence of a second critical point in water, proposed originally
in Ref.\cite{twoliquid} and later elaborated in several variants
(Refs.\cite{tanaka_nature}, \cite{tanaka_jcp}, \cite{dougherty04}).
According to this hypothesis, in the deeply supercooled region a second
critical point should exist, below which two kinds of water, high-density
liquid (HDL) and low-density liquid (LDL) are separated via a first-order
phase transition. Experimentally, however, it has not yet been possible to
access the deeply supercooled region and directly investigate its
properties. In the lack of direct evidence, the existence of two different
amorphous forms of ice has been used as an indirect support for this
hypothesis, assuming that HDA and LDA, apparently separated by a sharp
transition, are simply glassy forms of HDL and LDL\cite{pooletal}.

Recently several new experimental
results\cite{loerting,klotz,finney_vhda,tulk,koza,guthrie,mishima4,johari}
raised new questions about the nature of amorphous ices as well as about
their relation to the speculated second critical point of water. A new
amorphous form of ice has been reported\cite{loerting}, prepared by heating
HDA under pressure of 11 kbar to $T \sim 170$ K and cooling it back to
$T=80$ K; when recovered at ambient pressure it has a density of $\sim$
1.25 g/cm$^3$. It has been called very high-density amorphous ice (VHDA)
and characterized experimentally by neutron diffraction \cite{finney_vhda}.
More detailed structural measurements of VHDA ice were performed very
recently\cite{vhda_structure} and showed that the radial distribution
function (RDF) of VHDA is in fact more structured than that of HDA and LDA,
revealing the presence of an enhanced medium-range order. Other
experiments\cite{tulk, guthrie, koza, johari} focused on the HDA-LDA
transition induced by heating at ambient or low pressure.  In
Ref.\cite{tulk} it was shown that by heating HDA to temperatures
intermediate between 80 K and 110 K the sample can be trapped in an
apparent continuum of metastable structures between HDA and LDA. This
suggests that there might be no sharp transition between the two forms.  In
Ref.\cite{koza} the kinetics of the HDA to LDA transition was studied; this
revealed three stages of the transformation, consisting of the annealing of
HDA, followed by an accelerated transition to the LDA form and subsequent
annealing of the LDA. The experiment in Ref.\cite{mishima4}, while also
finding a continuum of HDA states, observed also a propagation of the
LDA-HDA interface, thus providing evidence in favor of a sharp transition
between the two forms.  Possible implications of the new experiments have
been discussed\cite{klug,soper,johari}.  Several review papers on amorphous
and supercooled water have also appeared recently, see
Refs.\cite{deben_stan},\cite{debenedetti},\cite{angell}.

Simulations can complement the experiment by providing access to shorter
time scales, not easily accessed in experiment. At the same time they can
provide detailed structural information which might not be easily extracted
from the experimental data. The basic phenomenology of amorphous ices has
been reproduced in earlier works\cite{klein,pooletal,okabe}. New
simulations have also been performed recently
\cite{yamada03},\cite{giovambattista03},\cite{brovchenko}, \cite{guissani},
\cite{giovambattista04},\cite{McBride04}.

Here we present the results of extensive constant-pressure MD simulations
of amorphous ice at temperatures 80 - 170 K and pressures up to 22.5 kbar.
We focus on the structure of the annealed form of amorphous ice and study
its evolution under increasing pressure. We identify the LDA and HDA
regimes, possibly separated by a transition. VHDA on the other hand does
not appear to be a new phase, but rather a particular high-pressure regime
of HDA.  In particular, we suggest that both new phenomena, the VHDA and
the continuum of metastable HDA densities at ambient pressure, originate
from the relationship between the density and the topology of the
hydrogen-bonded network of the HDA phase. A preliminary account of this
work has already been published in Ref.\cite{mdp_prl}. Here we present a
more detailed account of the results as well as new simulations and new
analysis.  The paper is organized as follows. In section
\ref{sec:model-technique} we present the model and details of our
simulation technique. In section \ref{sec:results} we present our results.
First we describe the protocol used to prepare various forms of amorphous
ice and then discuss in detail their properties, comparing with experiment
as well as with other simulations.  In section \ref{sec:analogy-silica} we
draw an analogy of another important tetrahedrally-bonded system, amorphous
silica.  Finally, in the last section \ref{sec:conclusions} we summarize
our findings and draw some conclusions concerning the relation between
amorphous ice and supercooled water.

\section{Model and technique}
\label{sec:model-technique}

Our tool is constant-pressure molecular dynamics (MD) simulation. We
employed the GROMACS code\cite{gromacs} using the Parrinello-Rahman
constant-pressure MD\cite{pr}, the Berendsen thermostat\cite{berendsen} and
a time step of 2 fs. Electrostatic interactions were treated by the
particle-mesh Ewald method\cite{pme}. An initial proton-disordered
configuration of I$_{\rm h}$ ice with 360 H$_2$O molecules and zero dipole
moment in an orthorhombic box was prepared using the Monte Carlo procedure
of Ref.\cite{buch}.

The interaction between water molecules was described by the classical
TIP4P model\cite{tip4p}. In previous simulations this was found to
reproduce well the transitions I$_{\rm h}$ -- HDA and LDA -- HDA, both
qualitatively and quantitatively\cite{klein,pooletal,okabe}. Although it is
known to systematically overestimate the experimental ice densities by
about 2 \%, the TIP4P model has recently been shown to be capable of
predicting qualitatively correctly the entire phase diagram of
water\cite{sanz04}.

\section{Results and discussion}
\label{sec:results}

\subsection{Preparation and annealing of amorphous ice}
\label{sec:prep-anne-amorph}

\begin{figure}[h]
  \includegraphics*[width=8.5cm]{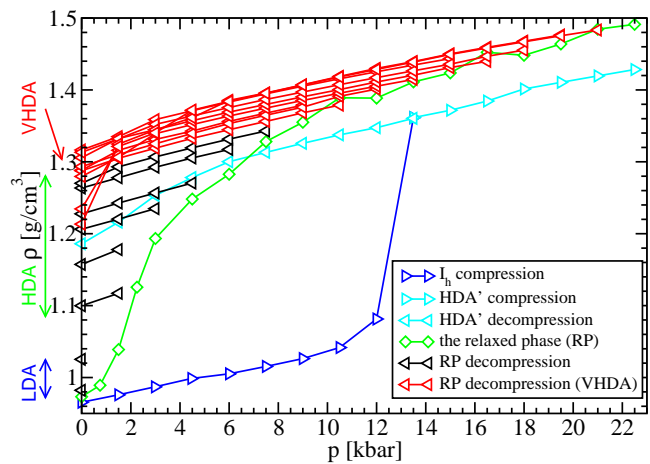}
\caption{Density vs.~pressure dependence at $T=80$~K for the various
  amorphous phases during compression/decompression.  The triangles point
  in the direction of pressure change, the lines are just a guide for the
  eye.}
\label{fig_ro_p}
\end{figure}

\begin{figure}[h]
  \includegraphics*[width=8.5cm]{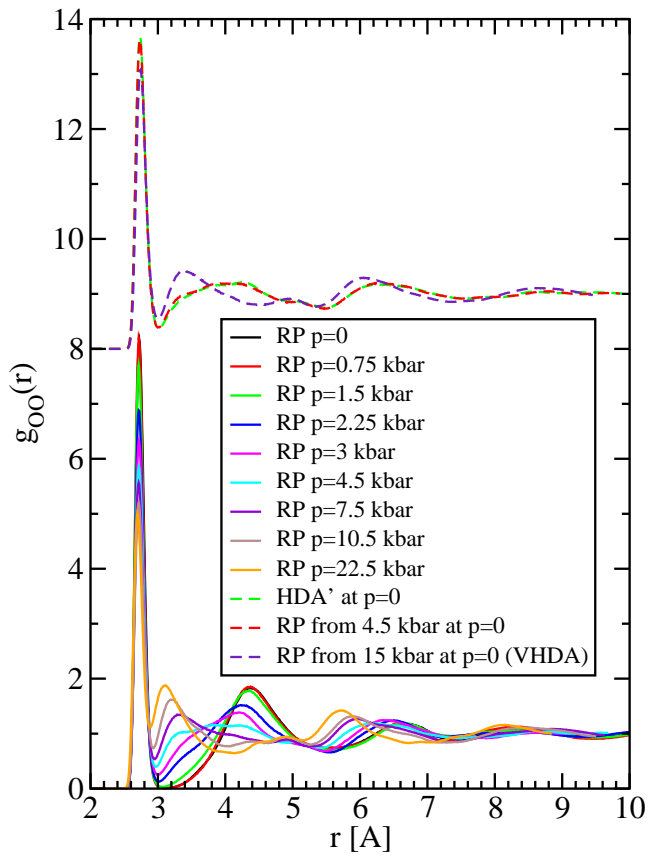}
\caption{Oxygen -- oxygen radial distribution function of various amorphous
  phases at $T=80$ K and $p = 0 - 22.5$ kbar. The curves in the upper part
  of the figure have been shifted by 8 for clarity.}
\label{fig_rdf}
\end{figure}

First we shall describe the simulation protocol that we applied. We started
by compressing the ice I$_{\rm h}$ at $T = 80$ K, increasing the pressure
in steps of 1.5 kbar.  At 13.5 kbar a sharp transition occurs and the
density increases by almost 30 \% to 1.37 g/cm$^3$ (Fig.\ref{fig_ro_p}).
The sample was then further compressed at $T=80$ K to 22.5 kbar and from 15
kbar decompressed to $p=0$. This particular HDA form obtained by
pressure-induced amorphization of I$_{\rm h}$ ice at $T = 80$ K and
subsequent compression or decompression, without any thermal treatment,
will be in the following denoted by HDA' (as-prepared HDA phase). During
decompression the HDA' density gradually decreased and at $p = 0$ reached
the value of 1.19 g/cm$^3$, close to the HDA experimental value of 1.17
g/cm$^3$\cite{mishima1}.  The radial distribution function (RDF) of HDA' at
$p=0$ is shown in Fig.\ref{fig_rdf}; it has a broad second peak between 3.3
and 4.6 \AA, very similar to that of HDA at $p=0$\cite{klotz}.  Inspired by
the experiments\cite{loerting} that led to the discovery of the VHDA we
decided to anneal the HDA' phase at each intermediate pressure between 22.5
kbar and zero (a similar treatment was applied in Ref.\cite{loerting} at $p
= 8.4, 11$ and 19 kbar) in order to search for possible new structures.
Annealing was performed by heating up to $T=170$ K and subsequent cooling
down to 80 K; the temperature was always changed in steps of 10 K.  The
phase obtained in this way will be called relaxed phase (RP). We note here
that in Ref.\cite{Tanaka2000} a slow molecular diffusion was observed in
the MD simulation of the TIP4P model at $T=173$ K; therefore at 170 K it
should be possible to equilibrate the system within accessible simulation
times. While in experiments a HDA' sample heated at an arbitrary pressure
might recrystallize\cite{pooletal,klug,klotz_recryst}, in the time scale of
a simulation this is not likely to happen. We are therefore restricted to
exploring the (metastable) disordered structures.  In order to check the
metastability of the RP phase upon decompression, the RP prepared at each
pressure was finally decompressed at $T=80$ K down to $p=0$, decreasing the
pressure in steps of $1.5$ kbar.

After each change of pressure or temperature the system was equilibrated
for 5 ns and observable quantities were averaged over 0.5 ns. An additional
equilibration for 25 -- 50 ns was performed during the annealing at the
highest temperature of $T=170$ K. We stress here that the long
equilibration times are indeed necessary in order to allow the system to
undergo structural changes; e.g at pressure $p=0.75$ kbar the equilibration
of the system at $T=170$ K takes about 20 ns.  A comprehensive summary of
the density vs. pressure dependence of I$_{\rm h}$, HDA', RP and
decompressed RP phases at $T=80$ K is presented in Fig.\ref{fig_ro_p} and
will be discussed in detail in the next section.
\begin{figure}[h]
  \includegraphics*[width=8.5cm]{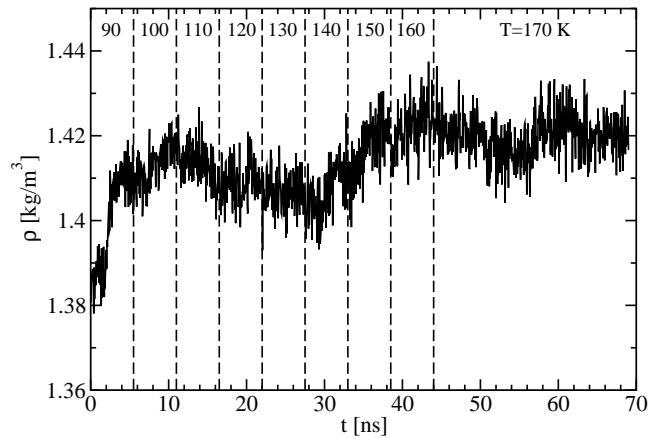}
\caption{Time dependence of the density during annealing of the HDA' ice at
  $p = 16.5$ kbar to $T=170$ K. Points where the temperature is increased
  by 10 K are marked by vertical dashed lines. The temperature (in K)
  during each time interval is shown in the top part of the figure.}
\label{fig_dens_heat}
\end{figure}

In Fig.\ref{fig_dens_heat} we show the relaxation of the density during
annealing of the HDA' ice at $p = 16.5$ kbar to $T=170$ K. It can be seen
that upon annealing at 90 and 100 K the density grows while between 110 and
130 K the sample undergoes a thermal expansion. At 140 K and 150 K the
density grows further but no substantial change is observed above $T=150$
K.

\begin{figure}[h]
  \includegraphics*[width=8.5cm]{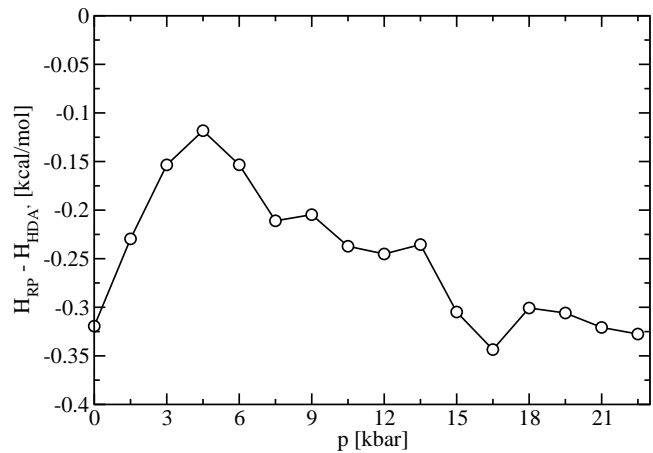}
\caption{Decrease of enthalpy upon annealing: difference between the
  enthalpy of the RP phase and that of the HDA' phase at T=80 K.}
\label{fig_enthalpy}
\end{figure}

We have verified that the enthalpy of the relaxed phase is lower than that
of HDA' ice at any pressure (Fig.\ref{fig_enthalpy}).  Assuming that the
entropy contribution to the Gibbs potential can be neglected at $T=80$ K
(and entropy differences between amorphous phases are anyway expected to be
small\cite{entropy}) this demonstrates that upon annealing at any pressure
HDA' ice irreversibly relaxes to a state with a lower free energy.  This a
posteriori justifies the necessity of annealing in order to reach a
metastable equilibrium corresponding at each pressure to a well-defined
thermodynamic phase. We note that the lowest difference between the
enthalpy of the HDA' phase and that of the RP is found at $p =$ 4.5 kbar,
suggesting that at the latter pressure HDA' is closest to its corresponding
relaxed amorphous form (see also next subsection).

\subsection{Evolution of the RP with increasing pressure}
\label{sec:evolution-rp-with}

In this section we analyze the properties of the RP and their dependence on
pressure, with the focus on structure. The density vs. pressure dependence
of the RP (Fig.~\ref{fig_ro_p}) can be considered as the equation of state
of amorphous ice.  We note first that the HDA' and RP curves cross at about
7 kbar; for lower pressure annealing results in lower density while for
higher pressures the system densifies.
 
At $p = 0$ the density after annealing reaches a value of 0.97 g/cm$^3$,
which agrees well with the experimental LDA value of 0.94 g/cm$^3$. The
remarkable feature of the RP curve is the narrow region between 1.5 and
2.25 kbar where the density increases by about 9 \%, from 1.04 g/cm$^3$ to
1.13 g/cm$^3$. Upon further increase of pressure the density grows fast and
reaches at $p=3$ kbar the value of 1.19 g/cm$^3$. Beyond that point the
density growth slows down progressively and at the highest pressure of 22.5
kbar the density reaches a value of 1.49 g/cm$^3$.

\begin{figure}[h]
  \includegraphics*[width=8.5cm]{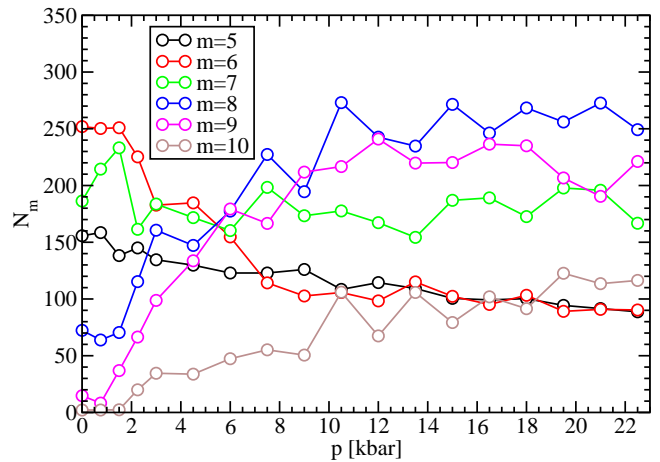}
\caption{Number of $n$-membered network rings in the RP as a function of
  pressure at $T=80$ K and $p = 0 - 22.5$ kbar. }
\label{fig_rings}
\end{figure}

The O-O RDF's of RP at different pressures are shown in Fig.\ref{fig_rdf}.
We also calculated the O-H RDF (not shown) for RP at $p=0, 2.25, 6$ and
$15$ kbar. Integrating between 1.5 and 2.25 \AA~we found at all pressures a
coordination number of 2, indicating a fully hydrogen-bonded network. In
order to characterize the evolution of the network topology we calculated
the ring statistics for RP at all pressures. This reveals information on
medium-range order that otherwise might not be easily extracted from the
RDF\cite{trachenko1,davila}. We applied the ring counting
algorithm\cite{yuan02} using the criterion from Ref.\cite{paolo04} to
identify the hydrogen bonds (we used for the O-O distance a cutoff
parameter of $r_{cut}=3.05 $ \AA~and a tolerance of $\Delta=0.2$ \AA).  Our
aim is to draw qualitative conclusions, so we did not try to bring down the
statistical error by repeating several times the quench from 170 K to 80 K
and averaging over the resulting structures.  We now discuss the evolution
of the RP at increasing pressure in terms of RDF and network ring
statistics (Fig.\ref{fig_rings}) and show that there are 3 distinct
regimes.

The RDF of the RP at $p = 0$ (Fig.\ref{fig_rdf}) exhibits at $r = 3.1$
\AA~a very deep minimum between the first and second shell and a
well-defined second shell peak at $r = 4.4$ \AA, very similar to that found
experimentally for LDA in Ref.\cite{finney_hlda}. The phase thus coincides
with the LDA as expected. The same is true at 0.75 kbar at a density of
0.99 g/cm$^3$, where the RDF within the second peak is practically
indistinguishable from the one at $p=0$, and only beyond 5 \AA~can small
differences be seen.  At 1.5 kbar the density increases to 1.04 g/cm$^3$
while the RDF becomes slightly different from that of LDA at $p=0$; the
very deep minimum between the first and second shell is still present. At
all these pressures the network is dominated by 6-membered rings.

The properties of the RP at $p=2.25$ kbar are rather different, correlating
with the sharp increase of density. The second shell peak of the RDF drops
and shifts to lower $r$ and at the same time the RDF grows substantially in
the region around $r = 3.3$ \AA, revealing the presence of interstitial
molecules\cite{klotz}. A dramatic change is seen in the ring statistics:
the number of 6-membered rings now starts to drop and at the same time the
number of 8-membered rings grows fast.  This behavior is compatible with a
transition from LDA to HDA occurring between 1.5 and 2.25 kbar.  Around
$p=4.5$ kbar the RDF develops a broad second peak between 3.2 and 4.4 \AA,
similar to that of HDA' at $p=0$ (Fig.\ref{fig_rdf}). This agrees with the
observation based on the enthalpy difference and shows that the HDA' is
indeed closest to the RP at this pressure.  Approaching $p \sim 10$ kbar
the ring statistics are definitely dominated by 8 and 9-membered rings; the
network has thus undergone a substantial reconstruction (see also section
\ref{sec:analogy-silica}). For $p > 10$ kbar, the ring statistics almost
stabilize, revealing that the reconstruction of the network is practically
completed; at the same time the density growth slows down further and the
compressibility approaches that of ice I$_{\rm h}$.  This indicates that
the increase of density due to the more efficient packing of the molecules
has at $p \sim 10$ kbar reached its limit at the value of $\rho_{lim} \sim
1.39$ g/cm$^3$, and further compression proceeds mainly by elastic
compression. In this limiting regime the RDF develops a pronounced second
peak at $r = 3.2$ \AA~(Fig.\ref{fig_rdf}) while the original second shell
peak at $r = 4.4$ \AA~disappears completely. In the subsection
\ref{sec:decompression-rp-p=0} we will identify this regime with the VHDA
form.

The above analysis provides quite a clear picture of the evolution of the
network topology when going from LDA to VHDA. Concerning this point, rather
contradictory opinions have been presented in the literature, based on the
indirect information provided by the detailed analysis of radial and
spatial distribution functions obtained from diffraction experiments and
the empirical potential structure refinement procedure\cite{epsr}. First we
note that there is no sign of any discontinuity upon evolution of the HDA
phase into the limiting VHDA regime. In Ref.\cite{finney_vhda} it was
speculated that there is no need to postulate any significant
reorganization of the network structure in moving between HDA and VHDA and
that they appear topologically isomorphous. Our results, however, show
clearly that the evolution of HDA between 2.25 kbar and 10 kbar must
inevitably involve a substantial reconstruction of the network.  On the
other hand, in Ref.\cite{klotz} it was suggested that HDA' under pressure
shows some characteristics of interpenetrating networks, similar to those
of high-pressure crystalline ices VII and VIII. We checked this feature in
the RP up to the highest pressure investigated, using the following
algorithm.  Starting from each molecule we considered a sphere with radius
$r_{cut} =$ 5 \AA~and tested whether the molecules within the sphere were
connected to the central one by a path containing no more than 4 hydrogen
bonds. We found that practically all molecules fulfilled this criterion;
this is clearly incompatible with the presence of interpenetrating
networks\cite{mdp_prl}. The VHDA structure can therefore be considered as
the upper limit to efficient amorphous packing of the molecules without
creating interpenetrating networks, as suggested in Ref.\cite{finney_vhda}.
Still, it is an interesting question whether amorphous ice with
interpenetrating networks can exist. Very recently, a study of VHDA was
performed\cite{saitta} in the region of densities ranging up to 1.9
g/cm$^3$, thus much higher than those studied here. Based on the analysis
of bond angle distributions, they showed that VHDA at very high densities
approaches a disordered close-packed structure, with local order more
similar to the fcc/hcp than to the bcc crystal. From this they concluded
that VHDA does not represent a disordered version of ice VII and therefore
does not have interpenetrating networks. The algorithm applied here in fact
provides direct evidence. We have also generated a sample of the RP at
$p=42$ kbar, with density $\rho = 1.61$ g/cm$^3$, and found that the ring
statistics were practically equal to the average of the RP in the interval
10 - 22.5 kbar. This shows that the network topology stabilizes already for
densities $\rho > \rho_{\rm lim} \sim 1.39$ g/cm$^3$, although the local
order converges only at much higher densities.

\subsection{Analysis of the shape of the rings}
\label{sec:analysis-shape-rings}

In order to obtain a deeper insight into the structural response to
pressure of the RP in the three regimes (LDA, HDA, VHDA), we also performed
a more detailed analysis of the network structure, separating the
contributions coming from different rings.

The shape of the rings has been characterized through the eigenvalues of
the inertia tensor. Given the three eigenvalues $I_1$, $I_2$ and $I_3$,
sorted by increasing magnitude, we define an elongation parameter
$\epsilon$ as $\epsilon=(I_2-I_1)/I_3$ and the asphericity ($\alpha$) as
the root mean square deviation of $I_i$ normalized to I$_3$. According to
the definitions, $\epsilon$ can vary from 0, for a circular ring or a
sphere, to 1 for a linear arrangement, while $\alpha$ is zero for a sphere,
0.236 for a circular ring and 0.58 for a linear arrangement. The
distribution of $\alpha$ for all the rings considered is peaked around
$\sim 0.2$ at $p=0$.  Increasing the pressure increases the spreads of the
distribution and shifts the peak toward higher $\alpha$. The tail of the
distribution never extends below 0.15, which means that the rings are
mainly planar structures.
\begin{figure}
  \includegraphics*[width=8.5cm]{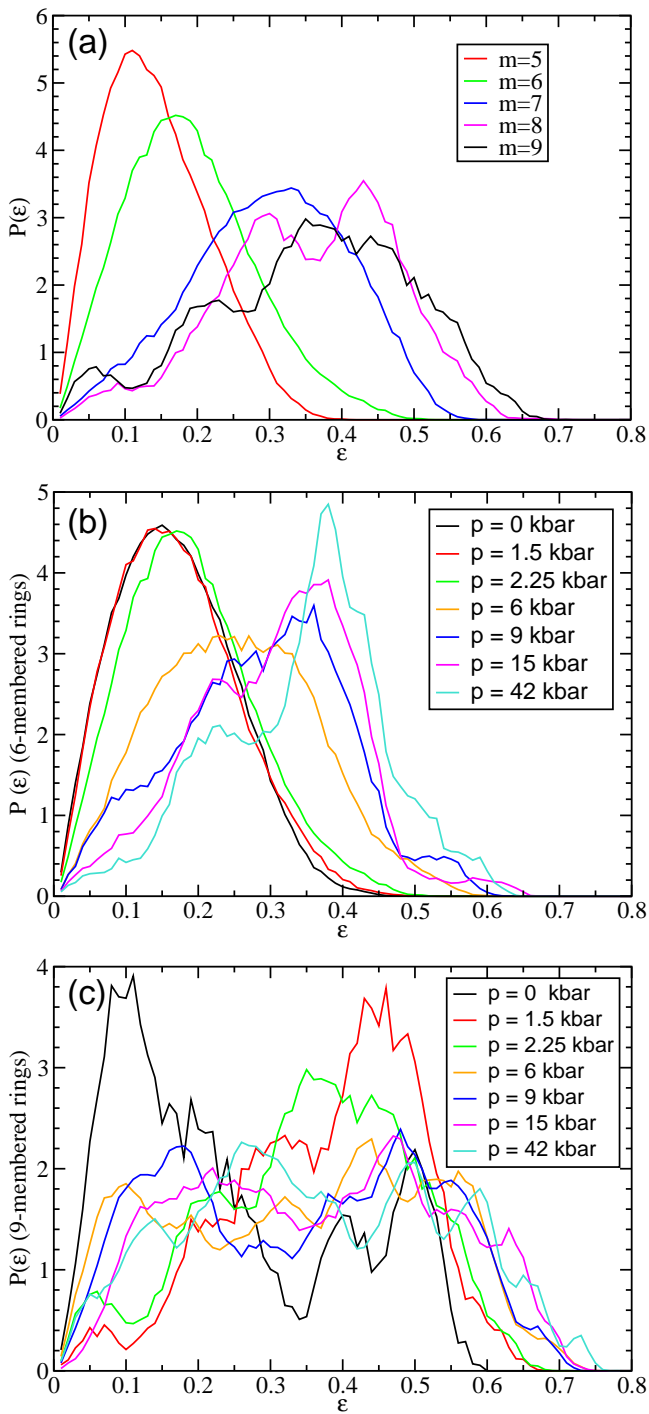}
\caption{(a) Probability distribution of the elongation parameter of the rings
  in HDA at 2.25 kbar. (b) and (c) P($\epsilon$) at several pressures from
  LDA to VHDA for six and nine-membered rings, respectively. We note that
  all the curves are normalized to one and therefore do not reflect the
  change of the total number of rings with a given size.}
\label{fig_elong}
\end{figure}

The probability distribution of the elongation parameter provides a clear
indication of the evolution of the shape of different rings as a function
of the pressure. In HDA at $p=2.25$ kbar (Fig.\ref{fig_elong}a) the
P($\epsilon$) shows that the water molecules in five and six-membered rings
arrange themselves in circular rings. The larger the ring the broader the
distribution, indicating that larger rings can arrange into elongated
structures without paying too much in terms of strain energy.
P($\epsilon$) at different pressures for six and nine-membered rings are
shown in Fig.\ref{fig_elong} panels (b) and (c), respectively. Six and
nine-membered rings, the quantity of which is most affected by
pressure-induced phase transitions, are shown as representative of the
behavior of small and large rings under pressure, respectively.  At low
pressures (up to 2.25 kbar) the shape of the small rings
(Fig.\ref{fig_elong}b) remains unchanged but when the transition to HDA
occurs their number rapidly decreases, and elongated large eight and
nine-membered rings are formed. In fact, already at $p=1.5$ kbar the main
peak of the $P(\epsilon)$ of nine-membered rings (Fig.\ref{fig_elong}c)
shifts from 0.1 to 0.48. The broad $P(\epsilon)$'s of large rings at higher
pressures show that such rings can easily adapt to any shape so as to
achieve a more efficient compaction.  In the HDA region the morphology of
small rings evolves toward more elongated shapes and their amount decreases
continuously.  When the onset pressure for the VHDA regime is reached, the
topology of the amorphous network stops changing and also the
$P(\epsilon)$'s stabilize. The residual small rings in VHDA are arranged
into elongated structures (the P($\epsilon$) is peaked at 0.4), whereas the
$P(\epsilon)$ of the nine-membered rings has no sharp peaks.  At this
regime the response to further compression consists in the deformation of
short-range structural features, such as the bond angle distributions, as
no more important rebonding is induced by increasing the pressure up to 42
kbar.
\begin{figure}
  \includegraphics*[width=8.5cm]{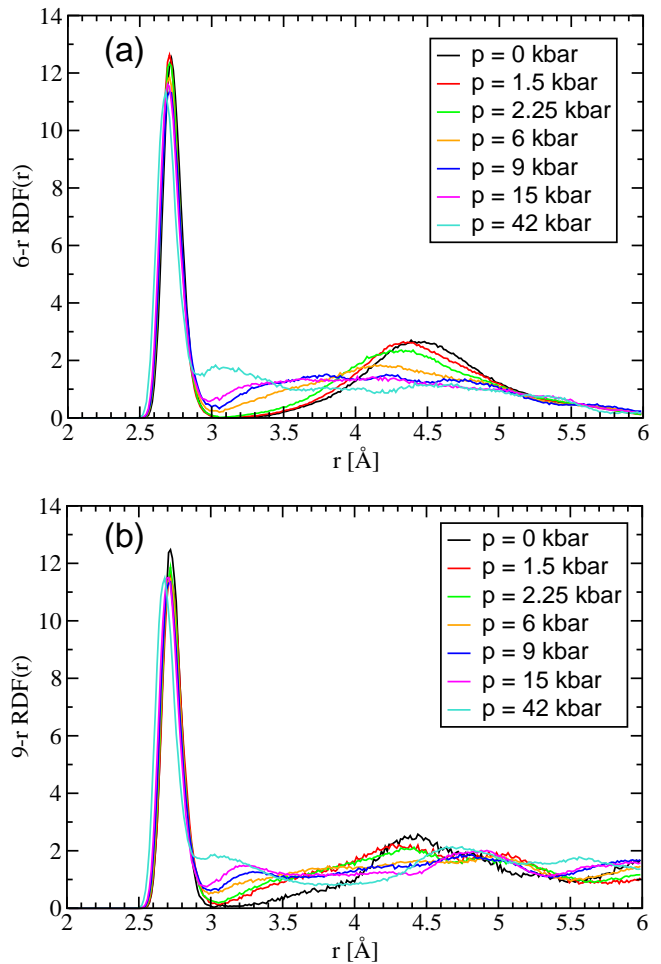}
\caption{Ring-restricted radial distribution function at several pressures 
for six (a) and nine-membered (b) rings.}
\label{fig_rRDF}
\end{figure}

We define a ring-restricted radial distribution function $n$-rRDF$(r)$ as
the probability of finding two atoms at a distance $r$ within the same
$n$-membered ring. This quantity allows us both to recognize the separate
contribution of different rings to the $g(r)$ and to characterize the
response of different rings to compression.  In Fig.\ref{fig_rRDF} the
oxygen-oxygen $n$-rRDF$(r)$ at different pressures are shown for six and
nine-membered rings.  The position of the first peak of both rRDF's is
unaffected by compression up to 15 kbar and even a compression to 42 kbar
induces a shift as small as 0.04 \AA. Low pressures up to 2.25 kbar do not
affect significantly the 6-rRDF, whereas in the same range of pressures
nine-membered rings already provide a sizable contribution to the
interstitial region between the first and the second peak of the $g(r)$. In
the region of stability of the HDA (6 kbar) six-membered rings are strained
and contribute to the interstitial region of the $g(r)$ through a
broadening and a shifting to the left of the second peak. On the other hand
the second peak of the 9-rRDF(r) spreads partly in the interstitial region
and gives rise to a shallow peak at 4.9 \AA. Such a peak becomes more
pronounced as the pressure increases and is a feature of VHDA
\cite{vhda_structure}. It is worth noting that even at 42 kbar the
six-membered rings do not provide a sharp contribution to this feature.  We
note that the rRDFs fully account for the interstitial peak in the $g(r)$
of VHDA, showing that this peak originates from contributions within the
same ring.  This constitutes independent evidence that there is no
formation of interpenetrating networks.

\subsection{Transition LDA-HDA}
\label{sec:transition-lda-hda}

The question of whether there is indeed a transition between LDA and HDA
and, if so, what its character is, is of great importance. From the ring
statistics shown in Fig.\ref{fig_rings} it can be seen that the response of
the number of rings to the applied pressure is rather different in the
regions below 1.5 and above 2.25 kbar.  Because of the limited accuracy of
the ring statistics calculated in the frozen states, we can conclude that
this behavior is compatible with the existence of a phase transition as
required by the second critical point scenario\cite{twoliquid}.
 
In order to determine whether the density of the RP changes discontinuously
between 1.5 and 2.25 kbar, we performed at $T=170$ K also a $\sim$ 30 ns
simulation at an intermediate pressure of 1.875 kbar. We did not observe
any metastability or hysteresis effects but instead rather large and slow
density fluctuations. Since our system is relatively small, such behavior
is compatible with a weak first order transition, occurring just below the
critical temperature, where the system oscillates over a low barrier
between two states. In principle it is, however, also possible that the
change of density with pressure is genuinely continuous, with a highly
compressible region between 1.5 and 2.25 kbar. In order to shed more light
on this important issue, it would be necessary to perform simulations with
larger systems, including several system sizes, and apply standard
finite-size scaling techniques\cite{finite-size-scaling}. It might also be
useful to perform free energy calculations, e.g. umbrella sampling, with a
suitable order parameter, similar to that performed in Ref.\cite{frenkel}.
Both techniques would, however, require considerable CPU resources to
achieve a good equilibration and sampling, since in the interesting region
the free energy surface is very rough, resulting in very long
autocorrelation times.

\subsection{Decompression of RP to $p=0$}
\label{sec:decompression-rp-p=0}

Most experimental data on amorphous ices have been gathered on samples
decompressed to ambient pressure. To our knowledge, there are no
experimental data for the RP under pressure to which we could directly
compare our results of the subsection \ref{sec:evolution-rp-with}. We
discuss now the interesting behavior of the RP upon decompression.
\begin{figure}[h]
  \includegraphics*[width=8.5cm]{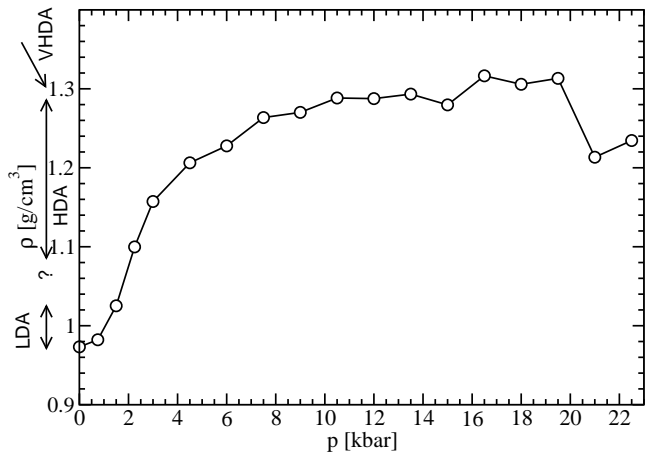}
\caption{Density of the RP after decompression at $T=80$ K from pressure $p$.}
\label{ro_RP_decompressed_vs_p}
\end{figure}

The density of the LDA phase from 0.75 kbar relaxes at $p=0$ to $\rho =
0.98$ g/cm$^3$, very close to the $\rho = 0.97$ g/cm$^3$ of LDA prepared at
$p=0$.  The LDA phase from 1.5 kbar on the other hand reaches at $p=0$ a
somewhat higher value of $\rho = 1.025$ g/cm$^3$, suggesting the
possibility that at $T=80$ K and $p=0$ there might not be a unique
structure of the LDA (in agreement with Refs.\cite{giovambattista03},
\cite{koza}) and this phase can actually span a narrow density interval.
In the pressure interval $p=2.25 - 10$ kbar, all decompression curves are
roughly parallel (Fig.\ref{fig_ro_p}) and the faster growing RP density
results upon decompression in an increasing density at $p=0$, spanning the
interval from 1.10 to 1.26 g/cm$^3$.  The picture changes remarkably for $p
> 10$ kbar.  Here, the slope of the RP curve becomes close to that of the
decompression curves which lie close to each other and initially almost
follow the RP curve.  Decompression from almost all pressures results at
$p=0$ in a narrow interval of densities around $\rho_{VHDA} \sim 1.29$
g/cm$^3$, corresponding to the decompression from the limiting density
$\rho_{lim}$.  For convenience, in Fig.\ref{ro_RP_decompressed_vs_p} we
show the dependence of the final $p=0$ density on the original pressure $p$
at which the RP was prepared, where the saturation can be clearly seen.
This agrees very well with the observation in Ref.\cite{loerting} where the
samples annealed at 11 and 19 kbar reached upon decompression the same VHDA
density of 1.25 g/cm$^3$; in fact, this was the reason why VHDA was
originally suspected to represent a new thermodynamic phase. The RDF of RP
decompressed from 15 kbar (Fig.\ref{fig_rdf}) is clearly similar to that of
VHDA recovered at $p=0$ in experiment\cite{finney_vhda}, showing the
presence of the distinct peak at $r=3.37$ \AA, very close to the first
shell peak. This allows us to identify this form as VHDA. The spectrum of
metastable states at $p=0$ (Fig.\ref{ro_RP_decompressed_vs_p}) is thus
compatible with the existence of a narrow LDA region and a broad continuum
of metastable HDA states with a density below that of VHDA (as found
experimentally in Ref.\cite{mishima4}).  We note that the density spectrum
of the HDA states extends both above and below that of the HDA', and there
is no reason to consider HDA' as a particular state representative of the
HDA phase. While the LDA states might be separated from the HDA ones by a
gap, it must certainly be much smaller than the density difference between
LDA and HDA' at $p=0$.

\begin{figure}[h]
  \includegraphics*[width=8.5cm]{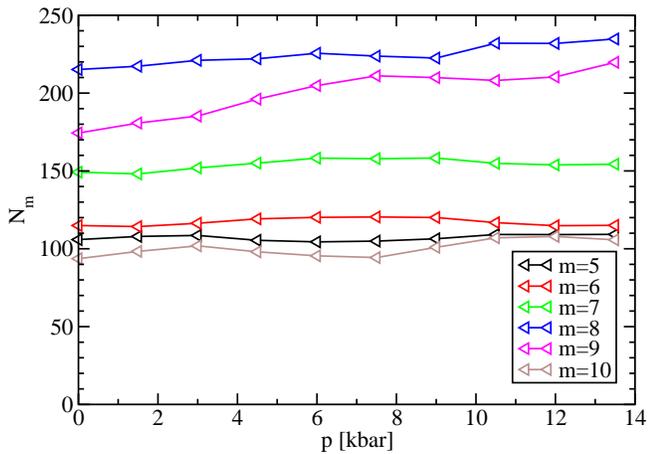}
\caption{Ring statistics during the decompression of RP at $T=80$ K from
  $p=13.5$ kbar to zero.}
\label{rings_decompression}
\end{figure}

We now suggest an explanation for the existence of an upper limit
$\rho_{VHDA}$ to the density of metastable HDA at $T=80$ K and $p=0$. In
the HDA phase with $\rho < \rho_{lim}$, the system with increasing pressure
achieves a better packing of molecules due to network reconstruction. This
must involve the breaking and remaking of bonds, which at any density
requires crossing a free energy barrier, and is only possible due to the
annealing at higher temperature, in our case 170 K. During cooling to 80 K,
the network topology becomes frozen. Upon subsequent decompression at 80 K,
the barriers cannot be crossed and therefore the system cannot relax to a
lower density via reconstruction of the network. This is illustrated in
Fig.\ref{rings_decompression} where the evolution of the ring statistics
upon decompression of the RP from $p=13.5$ kbar is shown; it can be seen
that no pronounced change in the network topology occurs. The largest
change is seen in the number of 9 and 8-membered rings, which decrease by
about 20 \% and 9\%, respectively, while the number of other rings stays
practically constant. The decompression thus proceeds dominantly via
relaxation of the elastic compression and that is why the HDA states with
$\rho < \rho_{lim}$, which have a variety of different topologies, relax
elastically to different states, spanning a range in density. On the other
hand, all HDA states with $\rho > \rho_{lim}$, which have almost the same
network topology, relax upon decompression to the same density
$\rho_{VHDA}$.  This accounts for the behavior observed in
Ref.\cite{loerting}, with no need to postulate VHDA to be a new phase, and
is also consistent with the fact that we did not observe any discontinuity
during the evolution from HDA to VHDA.  The origin of the behavior of VHDA
upon decompression therefore appears to be kinetic rather than
thermodynamic.

The two points from the highest pressures of 21 and 22.5 kbar reach upon
decompression a lower density, 1.21 and 1.23 g/cm$^3$, respectively.  We
believe that this is related to the fact that our decompression is
performed too fast, resulting in excessive elastic energy at $p=0$. This
may, in turn, allow some barriers to be crossed and enable a transition to
a density $\rho < \rho_{VHDA}$.

\begin{figure}[h]
  \includegraphics*[width=8.5cm]{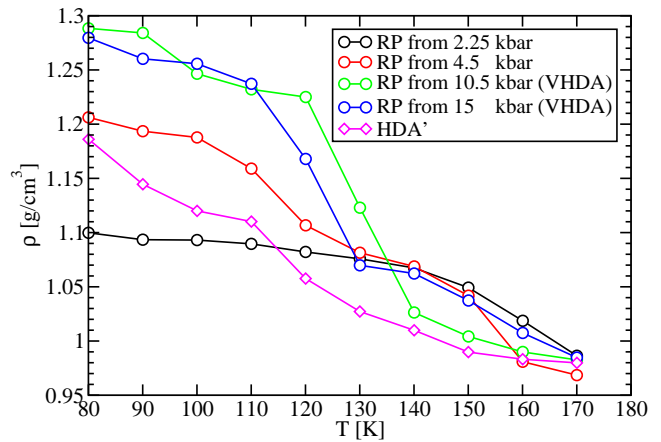}
\caption{Density as a function of temperature during heating of various
  decompressed RP phases as well as HDA' at $p=0$.}
\label{fig_hda_vhda_heat}
\end{figure}

We comment now on the experiments\cite{tulk,guthrie}, where HDA' was heated
to intermediate temperatures between 80 and 120 K and at each temperature
annealed for several hours. On this time scale every increase of
temperature resulted in an initial slow drop of the density which
afterwards gradually stabilized; an additional drop of the density could be
observed only by further increase of temperature. This behavior clearly
points to the fact that as the HDA phase approaches its low density limit,
the barriers increase and can be overcome only at a higher temperature. Our
picture of the structural evolution of the HDA phase is compatible with
these experimental findings. It is plausible to assume that the height of
the barriers is correlated to the amount of network reconstruction
necessary to change the volume, and is therefore related to the pressure
derivatives of the number of rings. As shown in section
\ref{sec:evolution-rp-with}, the reconstruction of the RP upon increasing
pressure is most dramatic at the low density limit of the HDA spectrum, and
with increasing density becomes gradually less pronounced, until it
practically vanishes for $\rho > \rho_{lim}$. We explicitly checked the
degree of metastability of HDA forms with different densities (RP
decompressed from several different pressures) by heating at $p=0$; the
results are shown in Fig.\ref{fig_hda_vhda_heat}.  The temperature was
increased in steps of 10 K, spending 5.5 ns at each temperature, and we
note that also here it would be useful to perform the heating several times
and average in order to improve the statistical error. Nevertheless, it can
be clearly seen that the most stable structure under heating is actually
the lowest density HDA (RP decompressed from 2.25 kbar) which undergoes
only a small drop in density up to 130 K. With increasing initial density
the samples start to relax at lower temperature, which confirms the above
relation between the network topology and barriers. It is also seen that
HDA' is the least stable of all the samples, as may be expected for an
insufficiently equilibrated phase possessing an excess free energy.  We
stress that at $T=170$ K all curves reach practically the same density,
lower than 0.99 g/cm$^3$, which also agrees with the experimental finding
in Refs.\cite{loerting},\cite{finney_vhda} that VHDA upon heating converts
to LDA. This is different from what found in Ref.\cite{guissani} where it
was argued that VHDA upon heating converts to a different form of LDA, with
a higher density of about 1.04 g/cm$^3$.

In Ref.\cite{guissani} various HDA forms were prepared by an alternative
procedure, consisting of the pressure-induced amorphization of cubic ice at
different temperatures ranging from 50 to 300 K.  It was concluded that the
HDA' produced by pressure-induced amorphization at liquid nitrogen
temperature and below represents a limiting form of HDA and thus the phase
spans a density interval between HDA' and VHDA.  This procedure does not
cover the part of the HDA spectrum that has a density below that of HDA'
and can exist at $p=0$ and $T=80$ K in metastable form and therefore we do
not consider the imperfectly equilibrated HDA' to be a limiting form of
HDA.

\begin{figure}[h]
  \includegraphics*[width=8.5cm]{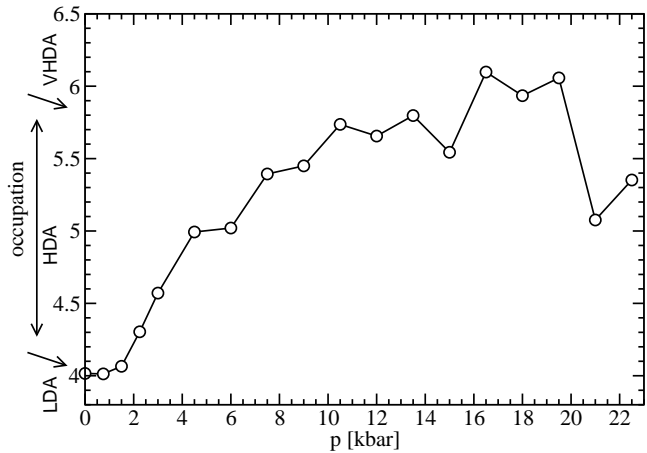}
\caption{Oxygen occupation number within 3.3 $\AA$ in the RP decompressed from
  pressure $p$ at $T=80$ K.}
\label{fig_occ_rp_decomp_p}
\end{figure}

In Ref.\cite{finney_vhda} the interstitial occupancy in HDA' and VHDA at
ambient pressure was calculated by integrating the O-O RDF between 2.3 and
3.3~\AA. The values of 5.0 and 5.8 were found, respectively, corresponding
to 50 \% or almost 100 \% occupancy of the ``lynch pin location''. It was
speculated that due to some unknown specific mechanism only these values
and no intermediate ones can be made stably at ambient pressure. We
calculated the same occupancy at $T=80$ K and $p=0$ in our HDA' as well as
in the RP decompressed from all pressures (Fig.\ref{fig_occ_rp_decomp_p}).
In HDA' we found a value of 4.9 while in the HDA branch of decompressed RP
we found an apparently continuous spectrum ranging from 4.3 (from 2.25
kbar) to about 6 (from pressures above 15 kbar). This again clearly shows
that while VHDA indeed represents a limiting structure, this is not the
case for HDA' whose occupancy close to the value of 5 can be regarded as
accidental.

\begin{figure}[h]
  \includegraphics*[width=8.5cm]{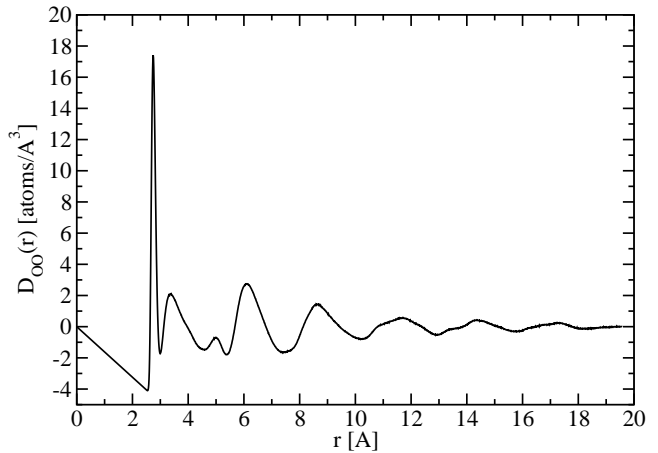}
\caption{Oxygen-oxygen radial distribution function $D_{OO}(r)$ for the
  VHDA prepared by annealing at $p=15$ kbar and decompressing to zero
  pressure.}
\label{fig_rdf_vhda}
\end{figure}

Recently, a neutron and X-ray diffraction study of the VHDA structure was
reported in Ref.\cite{vhda_structure}, showing the presence of at least
seven well-defined shells in the RDF of the VHDA, extending almost to the
distance of $\sim$ 20~\AA. This reveals the presence of an enhanced
medium-range order in the VHDA. In order to check this property, we also
prepared a bigger VHDA sample consisting of 2880 water molecules, allowing
us to calculate the RDF up to a distance of 20~\AA. We followed basically
the same protocol as for the 360-molecule sample and annealed the HDA' at
$p=15$ kbar, but using shorter simulation times.  The radial distribution
function $D_{OO}(r) = 4 \pi \rho r (g(r) -1)$ in the decompressed sample is
shown in Fig.\ref{fig_rdf_vhda}. Apart from the height of the first peak,
our result agrees well with the experimental one (Fig.2(b)) in
Ref.\cite{vhda_structure}, and also shows at least seven coordination
shells extending beyond $\sim$ 16~\AA.  The presence of such enhanced
medium-range order is likely to be related to the fact that the network
topology of VHDA is dominated by large rings. In Ref.\cite{vhda_structure}
the existence of a well-defined shell at 5~\AA~was pointed out; this is
also clearly present in our RDF and in subsection
\ref{sec:analysis-shape-rings} we have identified its origin in the
contribution of 9-membered rings (see Fig.\ref{fig_rRDF}(b)).

\section{Analogy to silica}
\label{sec:analogy-silica}

In spite of the different nature of the bonds between water molecules in
amorphous ice and between silicon and oxygen atoms in amorphous silica,
both systems consist of a continuous random network of corner-sharing
tetrahedra\cite{zach_cnr32} and in some windows of their phase diagrams
display analogous phenomenologies when pressure is applied.  Each
tetrahedral unit in silica is made of a four-fold coordinated silicon atom
in the center and four bridging oxygen atoms at the corners. The size
distribution of the primitive rings \cite{hobbs90,yuan02} is peaked at six
silicon atoms per ring and presents a sizable amount of four to ten-fold
rings \cite{hobbs90,yuan02,trachenko1}.  A high-density (HD) phase of
amorphous silica was discovered by Grimsditch \cite{grimsditch} about 20
years ago. It was shown that upon compression above 8 GPa a-SiO$_2$
undergoes a permanent densification which amounts to about 20$\%$ when the
system is released to ambient pressure and the transition to the HD phase
is accompanied by irreversible structural changes, observable by Brillouin
and Raman measurement \cite{grimsditch,hemley86}.

As in the case of amorphous ice, the nature of the phase transition is
still debated, since no discontinuous volume change is observed in
compression experiments \cite{grimsditch,hemley86,williams88} while
computer simulations do not clarify whether there is a kinetically hindered
first order transition \cite{lacks00} or a pressure window where there is a
balance between two densification mechanisms \cite{trachenko1,trachenko2}.
In fact three different microscopic mechanisms cooperate in accommodating
the amorphous silica network in a smaller volume
\cite{huang1,huang2,trachenko1,trachenko2} and make the phase transition
between LDA and HDA apparently smoother than in amorphous ice. In the low
pressure regime ($\simeq$3 GPa according to
Refs.\cite{trachenko1,trachenko2}) the volume diminishes only by the
buckling of the network, which results in a shift toward smaller values of
the Si-O-Si bond angle distribution. In this process the tetrahedral units
are not deformed and no bonds are broken.  At higher pressures the buckling
mechanism is supplemented by a substantial rebonding in the network, which
mainly affects the medium-range order features: the local tetrahedral order
is preserved but the ring-size distribution broadens and its peak shifts to
larger rings \cite{huang2}. In the response of silica to pressure the
buckling and the rebonding mechanisms correspond to elastic and plastic
regimes, respectively, as observed by D\'{a}vila {\it et al.} in MD
simulations \cite{davila}.  On the other hand, in both regimes short-range
structural properties such as the Si-O-Si and the O-Si-O angles vary
continuously.  In addition, coordination defects may be formed and
contribute to densification at even higher pressures (e.g.  $>5$ GPa in
Ref.\cite{trachenko1}).  In ice the nature of the hydrogen bond inhibits
this latter mechanism, as no more than four hydrogen bonds per water
molecule can be formed: in fact the average coordination number of the RP
is $\sim$4 at all the explored pressures.  Consequently, when the limit for
the topological densification is reached, amorphous ice turns back to an
elastic regime (VHDA).

The increasing of the characteristic size of the rings upon densification
is a general feature of both amorphous
\cite{hobbs90,stixrude_amorph,grimsditch_b2o3} and crystalline
\cite{marians_cryst} tetrahedral networks. Among the tetrahedrally
coordinated crystalline silica polymorphs, the lower density forms
(cristobalite and tridimite) consist of six-membered rings only.  In the
denser silica crystals the average size of the rings increases accordingly
to the density.  For example coesite, which is 30$\%$ denser than
$\alpha$-cristobalite has an average ring size of ten and contains rings of
size up to twelve.  It is worth noting that unlike the HD crystalline
phases of ice, it is impossible to form silica polymorphs with
interpenetrating networks, because of sterical interactions.

\section{Conclusions}
\label{sec:conclusions}

Upon increase of pressure, relaxed amorphous ice undergoes a pronounced
change of structure, from LDA at $p=0$ to VHDA at $p > 10$ kbar. During
this transformation, there is possibly a discontinuity between LDA and HDA,
although from our simulations performed on a relatively small system we
cannot distinguish between a weak first-order transition and a continuous
change.  Nevertheless, we can clearly distinguish the LDA and HDA regimes.
This identification is based on the existence of a transition region
characterized by a rapid change of density, on the presence of interstitial
molecules and the behavior of the ring statistics. It is important to note
that the main part of the overall change of the network topology does not
occur during the LDA-HDA transition (similar conclusion was also made in
Ref.\cite{koza}) but rather within the HDA phase, between $p \sim$ 2 and 10
kbar.  Concerning the as-prepared HDA', initially believed to be the only
possible form of HDA, it does not seem to have any special importance and
represents just one particular structure within the HDA megabasin. It is
not in equilibrium even within the space of amorphous structures and its
properties are determined by the conditions of preparation\cite{guissani}.
As shown above, when prepared at $T=80$ K, HDA' is rather close to RP at
4.5 kbar.  The HDA megabasin includes a broad range of structures with
different local and medium-range order and also spans a broad interval of
densities. On the high-density side, the onset of the VHDA regime marks the
limit to which the densification can be pushed by adapting the network
topology, without creating interpenetrating networks.  The low density
limit of HDA stability is more difficult to assess with precision. However,
as shown experimentally by Refs.\cite{tulk},\cite{mishima4},\cite{koza} and
also in our simulations, the HDA region reaches substantially below HDA'
density.

From the above it is clear that many forms of HDA exist and are metastable
at $T=80$ K upon decompression to $p=0$. The important question then is, to
what extent does this affect the use of the LDA/HDA phenomenology as
support for the conjecture for the second critical point in water. In
Ref.\cite{johari} it is claimed that since no unique HDA exists, it is
difficult to justify the conjecture of a second critical point for water.
We do not actually think that this is necessarily the case.  In our
picture, the variety of HDA ices which exist as metastable forms at $p=0$
corresponds to the variety of topologically different HDL at different
pressures. These various HDA forms cannot interconvert upon decompression
at $T=80$ K because this would involve rebonding and require overcoming
free energy barriers. In Ref.\cite{giovambattista04} it was suggested that
VHDA is a better candidate for the glassy phase continuous with the HDL,
rather than HDA'. Since we reserve the use of VHDA for the particular
high-pressure regime, we can say that all forms of the HDA branch of the RP
appear to be equally good candidates for the glassy phase that is the
putative continuation of the HDL. Consequently, the large density variation
of HDA ice at $p=0$ might reflect the existence of a region of high
compressibility of the supercooled HDL just below the critical
point\cite{yamada03}.  As suggested in
Refs.\cite{koza},\cite{giovambattista03}, various forms of LDA ice may also
exist at $T=80$ K and $p=0$, but their density variation is likely to be
much smaller.  A sharp transition may well exist between the upper limit of
LDA and lower limit of HDA continua, as suggested by the experiments in
Ref.\cite{koza}.  To shed more light on this issue, simulations on larger
systems combined with finite-size scaling techniques, as well as free
energy calculations could be helpful.

In the discussion of the polyamorphism of ice its analogy with the
properties of silica glasses can be useful, although attention should also
be paid to important differences.

We should like to acknowledge stimulating discussions with V.~Buch,
D.~Chandler, D.~D.~Klug, M.~M.~Koza and J.~S.~Tse.

\bibliography{paper}

\begin{thebibliography}{62}
\expandafter\ifx\csname natexlab\endcsname\relax\def\natexlab#1{#1}\fi
\expandafter\ifx\csname bibnamefont\endcsname\relax
  \def\bibnamefont#1{#1}\fi
\expandafter\ifx\csname bibfnamefont\endcsname\relax
  \def\bibfnamefont#1{#1}\fi
\expandafter\ifx\csname citenamefont\endcsname\relax
  \def\citenamefont#1{#1}\fi
\expandafter\ifx\csname url\endcsname\relax
  \def\url#1{\texttt{#1}}\fi
\expandafter\ifx\csname urlprefix\endcsname\relax\def\urlprefix{URL }\fi
\providecommand{\bibinfo}[2]{#2}
\providecommand{\eprint}[2][]{\url{#2}}

\bibitem[{\citenamefont{Mishima et~al.}(1984)\citenamefont{Mishima, Calvert,
  and Whalley}}]{mishima1}
\bibinfo{author}{\bibfnamefont{O.}~\bibnamefont{Mishima}},
  \bibinfo{author}{\bibfnamefont{L.~D.} \bibnamefont{Calvert}},
  \bibnamefont{and} \bibinfo{author}{\bibfnamefont{E.}~\bibnamefont{Whalley}},
  \bibinfo{journal}{Nature (London)} \textbf{\bibinfo{volume}{310}},
  \bibinfo{pages}{393} (\bibinfo{year}{1984}).

\bibitem[{\citenamefont{Mishima et~al.}(1985)\citenamefont{Mishima, Calvert,
  and Whalley}}]{mishima2}
\bibinfo{author}{\bibfnamefont{O.}~\bibnamefont{Mishima}},
  \bibinfo{author}{\bibfnamefont{L.~D.} \bibnamefont{Calvert}},
  \bibnamefont{and} \bibinfo{author}{\bibfnamefont{E.}~\bibnamefont{Whalley}},
  \bibinfo{journal}{Nature (London)} \textbf{\bibinfo{volume}{314}},
  \bibinfo{pages}{76} (\bibinfo{year}{1985}).

\bibitem[{\citenamefont{Mishima}(1994)}]{mishima3}
\bibinfo{author}{\bibfnamefont{O.}~\bibnamefont{Mishima}}, \bibinfo{journal}{J.
  Chem. Phys.} \textbf{\bibinfo{volume}{100}}, \bibinfo{pages}{5910}
  (\bibinfo{year}{1994}).

\bibitem[{\citenamefont{Poole et~al.}(1992)\citenamefont{Poole, Sciortino,
  Essmann, and Stanley}}]{twoliquid}
\bibinfo{author}{\bibfnamefont{P.~H.} \bibnamefont{Poole}},
  \bibinfo{author}{\bibfnamefont{F.}~\bibnamefont{Sciortino}},
  \bibinfo{author}{\bibfnamefont{U.}~\bibnamefont{Essmann}}, \bibnamefont{and}
  \bibinfo{author}{\bibfnamefont{H.~E.} \bibnamefont{Stanley}},
  \bibinfo{journal}{Nature} \textbf{\bibinfo{volume}{360}},
  \bibinfo{pages}{324} (\bibinfo{year}{1992}).

\bibitem[{\citenamefont{Tanaka}(1996{\natexlab{a}})}]{tanaka_nature}
\bibinfo{author}{\bibfnamefont{H.}~\bibnamefont{Tanaka}},
  \bibinfo{journal}{Nature} \textbf{\bibinfo{volume}{380}}, \bibinfo{pages}{328
  } (\bibinfo{year}{1996}{\natexlab{a}}).

\bibitem[{\citenamefont{Tanaka}(1996{\natexlab{b}})}]{tanaka_jcp}
\bibinfo{author}{\bibfnamefont{H.}~\bibnamefont{Tanaka}}, \bibinfo{journal}{J.
  Chem. Phys.} \textbf{\bibinfo{volume}{105}}, \bibinfo{pages}{5099}
  (\bibinfo{year}{1996}{\natexlab{b}}).

\bibitem[{\citenamefont{Dougherty}(2004)}]{dougherty04}
\bibinfo{author}{\bibfnamefont{R.~C.} \bibnamefont{Dougherty}},
  \bibinfo{journal}{Chem. Phys.} \textbf{\bibinfo{volume}{298}},
  \bibinfo{pages}{307} (\bibinfo{year}{2004}).

\bibitem[{\citenamefont{Poole et~al.}(1993)\citenamefont{Poole, Essmann,
  Sciortino, and Stanley}}]{pooletal}
\bibinfo{author}{\bibfnamefont{P.~H.} \bibnamefont{Poole}},
  \bibinfo{author}{\bibfnamefont{U.}~\bibnamefont{Essmann}},
  \bibinfo{author}{\bibfnamefont{F.}~\bibnamefont{Sciortino}},
  \bibnamefont{and} \bibinfo{author}{\bibfnamefont{H.~E.}
  \bibnamefont{Stanley}}, \bibinfo{journal}{Phys. Rev. E}
  \textbf{\bibinfo{volume}{48}}, \bibinfo{pages}{4605} (\bibinfo{year}{1993}).

\bibitem[{\citenamefont{Loerting et~al.}(2001)\citenamefont{Loerting, Salzmann,
  Kohl, Mayer, and Hallbrucker}}]{loerting}
\bibinfo{author}{\bibfnamefont{T.}~\bibnamefont{Loerting}},
  \bibinfo{author}{\bibfnamefont{C.}~\bibnamefont{Salzmann}},
  \bibinfo{author}{\bibfnamefont{I.}~\bibnamefont{Kohl}},
  \bibinfo{author}{\bibfnamefont{E.}~\bibnamefont{Mayer}}, \bibnamefont{and}
  \bibinfo{author}{\bibfnamefont{A.}~\bibnamefont{Hallbrucker}},
  \bibinfo{journal}{Phys. Chem. Chem. Phys.} \textbf{\bibinfo{volume}{3}},
  \bibinfo{pages}{5355} (\bibinfo{year}{2001}).

\bibitem[{\citenamefont{Klotz et~al.}(2002)\citenamefont{Klotz, Hamel, Loveday,
  Nelmes, Guthrie, and Soper}}]{klotz}
\bibinfo{author}{\bibfnamefont{S.}~\bibnamefont{Klotz}},
  \bibinfo{author}{\bibfnamefont{G.}~\bibnamefont{Hamel}},
  \bibinfo{author}{\bibfnamefont{J.~S.} \bibnamefont{Loveday}},
  \bibinfo{author}{\bibfnamefont{R.~J.} \bibnamefont{Nelmes}},
  \bibinfo{author}{\bibfnamefont{M.}~\bibnamefont{Guthrie}}, \bibnamefont{and}
  \bibinfo{author}{\bibfnamefont{A.~K.} \bibnamefont{Soper}},
  \bibinfo{journal}{Phys. Rev. Lett.} \textbf{\bibinfo{volume}{89}},
  \bibinfo{pages}{285502} (\bibinfo{year}{2002}).

\bibitem[{\citenamefont{Finney et~al.}(2002{\natexlab{a}})\citenamefont{Finney,
  Bowron, Soper, Loerting, Mayer, and Hallbrucker}}]{finney_vhda}
\bibinfo{author}{\bibfnamefont{J.~L.} \bibnamefont{Finney}},
  \bibinfo{author}{\bibfnamefont{D.~T.} \bibnamefont{Bowron}},
  \bibinfo{author}{\bibfnamefont{A.~K.} \bibnamefont{Soper}},
  \bibinfo{author}{\bibfnamefont{T.}~\bibnamefont{Loerting}},
  \bibinfo{author}{\bibfnamefont{E.}~\bibnamefont{Mayer}}, \bibnamefont{and}
  \bibinfo{author}{\bibfnamefont{A.}~\bibnamefont{Hallbrucker}},
  \bibinfo{journal}{Phys. Rev. Lett.} \textbf{\bibinfo{volume}{89}},
  \bibinfo{pages}{205503} (\bibinfo{year}{2002}{\natexlab{a}}).

\bibitem[{\citenamefont{Tulk et~al.}(2002)\citenamefont{Tulk, Benmore, Urquidi,
  Klug, Neuefeind, Tomberli, and Egelstaff}}]{tulk}
\bibinfo{author}{\bibfnamefont{C.~A.} \bibnamefont{Tulk}},
  \bibinfo{author}{\bibfnamefont{C.~J.} \bibnamefont{Benmore}},
  \bibinfo{author}{\bibfnamefont{J.}~\bibnamefont{Urquidi}},
  \bibinfo{author}{\bibfnamefont{D.~D.} \bibnamefont{Klug}},
  \bibinfo{author}{\bibfnamefont{J.}~\bibnamefont{Neuefeind}},
  \bibinfo{author}{\bibfnamefont{B.}~\bibnamefont{Tomberli}}, \bibnamefont{and}
  \bibinfo{author}{\bibfnamefont{P.~A.} \bibnamefont{Egelstaff}},
  \bibinfo{journal}{Science} \textbf{\bibinfo{volume}{297}},
  \bibinfo{pages}{1320} (\bibinfo{year}{2002}).

\bibitem[{\citenamefont{Koza et~al.}(2003)\citenamefont{Koza, Schober, Fischer,
  Hansen, and Fujara}}]{koza}
\bibinfo{author}{\bibfnamefont{M.~M.} \bibnamefont{Koza}},
  \bibinfo{author}{\bibfnamefont{H.}~\bibnamefont{Schober}},
  \bibinfo{author}{\bibfnamefont{H.~E.} \bibnamefont{Fischer}},
  \bibinfo{author}{\bibfnamefont{T.}~\bibnamefont{Hansen}}, \bibnamefont{and}
  \bibinfo{author}{\bibfnamefont{F.}~\bibnamefont{Fujara}},
  \bibinfo{journal}{J. Phys.: Condens. Matter} \textbf{\bibinfo{volume}{15}},
  \bibinfo{pages}{321} (\bibinfo{year}{2003}).

\bibitem[{\citenamefont{Guthrie et~al.}(2003)\citenamefont{Guthrie, Urquidi,
  Tulk, Benmore, Klug, and Neuefeind}}]{guthrie}
\bibinfo{author}{\bibfnamefont{M.}~\bibnamefont{Guthrie}},
  \bibinfo{author}{\bibfnamefont{J.}~\bibnamefont{Urquidi}},
  \bibinfo{author}{\bibfnamefont{C.~A.} \bibnamefont{Tulk}},
  \bibinfo{author}{\bibfnamefont{C.~J.} \bibnamefont{Benmore}},
  \bibinfo{author}{\bibfnamefont{D.~D.} \bibnamefont{Klug}}, \bibnamefont{and}
  \bibinfo{author}{\bibfnamefont{J.}~\bibnamefont{Neuefeind}},
  \bibinfo{journal}{Phys. Rev. B} \textbf{\bibinfo{volume}{68}},
  \bibinfo{pages}{184110} (\bibinfo{year}{2003}).

\bibitem[{\citenamefont{Mishima and Suzuki}(2002)}]{mishima4}
\bibinfo{author}{\bibfnamefont{O.}~\bibnamefont{Mishima}} \bibnamefont{and}
  \bibinfo{author}{\bibfnamefont{Y.}~\bibnamefont{Suzuki}},
  \bibinfo{journal}{Nature} \textbf{\bibinfo{volume}{419}},
  \bibinfo{pages}{599} (\bibinfo{year}{2002}).

\bibitem[{\citenamefont{Johari and Andersson}(2004)}]{johari}
\bibinfo{author}{\bibfnamefont{G.}~\bibnamefont{Johari}} \bibnamefont{and}
  \bibinfo{author}{\bibfnamefont{O.}~\bibnamefont{Andersson}},
  \bibinfo{journal}{J. Chem. Phys.} \textbf{\bibinfo{volume}{120}},
  \bibinfo{pages}{6207} (\bibinfo{year}{2004}).

\bibitem[{\citenamefont{Guthrie et~al.}(2004)\citenamefont{Guthrie, Tulk,
  Benmore, and Klug}}]{vhda_structure}
\bibinfo{author}{\bibfnamefont{M.}~\bibnamefont{Guthrie}},
  \bibinfo{author}{\bibfnamefont{C.~A.} \bibnamefont{Tulk}},
  \bibinfo{author}{\bibfnamefont{C.~J.} \bibnamefont{Benmore}},
  \bibnamefont{and} \bibinfo{author}{\bibfnamefont{D.~D.} \bibnamefont{Klug}},
  \bibinfo{journal}{Chem. Phys. Lett.} \textbf{\bibinfo{volume}{397}},
  \bibinfo{pages}{335} (\bibinfo{year}{2004}).

\bibitem[{\citenamefont{Klug}(2002)}]{klug}
\bibinfo{author}{\bibfnamefont{D.~D.} \bibnamefont{Klug}},
  \bibinfo{journal}{Nature} \textbf{\bibinfo{volume}{420}},
  \bibinfo{pages}{749} (\bibinfo{year}{2002}).

\bibitem[{\citenamefont{Soper}(2002)}]{soper}
\bibinfo{author}{\bibfnamefont{A.~K.} \bibnamefont{Soper}},
  \bibinfo{journal}{Science} \textbf{\bibinfo{volume}{297}},
  \bibinfo{pages}{1288} (\bibinfo{year}{2002}).

\bibitem[{\citenamefont{Debenedetti and Stanley}(2003)}]{deben_stan}
\bibinfo{author}{\bibfnamefont{P.}~\bibnamefont{Debenedetti}} \bibnamefont{and}
  \bibinfo{author}{\bibfnamefont{H.}~\bibnamefont{Stanley}},
  \bibinfo{journal}{Phys. Today} \textbf{\bibinfo{volume}{56}},
  \bibinfo{pages}{40} (\bibinfo{year}{2003}).

\bibitem[{\citenamefont{Debenedetti}(2003)}]{debenedetti}
\bibinfo{author}{\bibfnamefont{P.}~\bibnamefont{Debenedetti}},
  \bibinfo{journal}{J. Phys.: Condens. Matter} \textbf{\bibinfo{volume}{15}},
  \bibinfo{pages}{R1669} (\bibinfo{year}{2003}).

\bibitem[{\citenamefont{Angell}(2004)}]{angell}
\bibinfo{author}{\bibfnamefont{C.}~\bibnamefont{Angell}},
  \bibinfo{journal}{Annu. Rev. Phys. Chem.} \textbf{\bibinfo{volume}{55}},
  \bibinfo{pages}{559} (\bibinfo{year}{2004}).

\bibitem[{\citenamefont{Tse and Klein}(1987)}]{klein}
\bibinfo{author}{\bibfnamefont{J.~S.} \bibnamefont{Tse}} \bibnamefont{and}
  \bibinfo{author}{\bibfnamefont{M.~L.} \bibnamefont{Klein}},
  \bibinfo{journal}{Phys. Rev. Lett.} \textbf{\bibinfo{volume}{58}},
  \bibinfo{pages}{1672} (\bibinfo{year}{1987}).

\bibitem[{\citenamefont{Okabe et~al.}(1996)\citenamefont{Okabe, Tanaka, and
  Nakanishi}}]{okabe}
\bibinfo{author}{\bibfnamefont{I.}~\bibnamefont{Okabe}},
  \bibinfo{author}{\bibfnamefont{H.}~\bibnamefont{Tanaka}}, \bibnamefont{and}
  \bibinfo{author}{\bibfnamefont{K.}~\bibnamefont{Nakanishi}},
  \bibinfo{journal}{Phys. Rev. E} \textbf{\bibinfo{volume}{53}},
  \bibinfo{pages}{2638} (\bibinfo{year}{1996}).

\bibitem[{\citenamefont{Yamada et~al.}(2003)\citenamefont{Yamada, Stanley, and
  Sciortino}}]{yamada03}
\bibinfo{author}{\bibfnamefont{M.}~\bibnamefont{Yamada}},
  \bibinfo{author}{\bibfnamefont{H.~E.} \bibnamefont{Stanley}},
  \bibnamefont{and}
  \bibinfo{author}{\bibfnamefont{F.}~\bibnamefont{Sciortino}},
  \bibinfo{journal}{Phys. Rev. E} \textbf{\bibinfo{volume}{67}},
  \bibinfo{pages}{010202(R)} (\bibinfo{year}{2003}).

\bibitem[{\citenamefont{Giovambattista
  et~al.}(2003)\citenamefont{Giovambattista, Stanley, and
  Sciortino}}]{giovambattista03}
\bibinfo{author}{\bibfnamefont{N.}~\bibnamefont{Giovambattista}},
  \bibinfo{author}{\bibfnamefont{H.~E.} \bibnamefont{Stanley}},
  \bibnamefont{and}
  \bibinfo{author}{\bibfnamefont{F.}~\bibnamefont{Sciortino}},
  \bibinfo{journal}{Phys. Rev. Lett.} \textbf{\bibinfo{volume}{91}},
  \bibinfo{pages}{115504} (\bibinfo{year}{2003}).

\bibitem[{\citenamefont{Brovchenko et~al.}(2003)\citenamefont{Brovchenko,
  Geiger, and Oleinikova}}]{brovchenko}
\bibinfo{author}{\bibfnamefont{I.}~\bibnamefont{Brovchenko}},
  \bibinfo{author}{\bibfnamefont{A.}~\bibnamefont{Geiger}}, \bibnamefont{and}
  \bibinfo{author}{\bibfnamefont{A.}~\bibnamefont{Oleinikova}},
  \bibinfo{journal}{J. Chem. Phys.} \textbf{\bibinfo{volume}{118}},
  \bibinfo{pages}{9473} (\bibinfo{year}{2003}).

\bibitem[{\citenamefont{Guillot and Guissani}(2003)}]{guissani}
\bibinfo{author}{\bibfnamefont{B.}~\bibnamefont{Guillot}} \bibnamefont{and}
  \bibinfo{author}{\bibfnamefont{Y.}~\bibnamefont{Guissani}},
  \bibinfo{journal}{J. Chem. Phys.} \textbf{\bibinfo{volume}{119}},
  \bibinfo{pages}{11740} (\bibinfo{year}{2003}).

\bibitem[{\citenamefont{Giovambattista
  et~al.}(2004)\citenamefont{Giovambattista, Stanley, and
  Sciortino}}]{giovambattista04}
\bibinfo{author}{\bibfnamefont{N.}~\bibnamefont{Giovambattista}},
  \bibinfo{author}{\bibfnamefont{H.~E.} \bibnamefont{Stanley}},
  \bibnamefont{and} \bibinfo{author}{\bibfnamefont{F.}~\bibnamefont{Sciortino}}
  (\bibinfo{year}{2004}), \bibinfo{note}{cond-mat/0403365 preprint}.

\bibitem[{\citenamefont{McBride et~al.}(2004)\citenamefont{McBride, Vega, Sanz,
  and Abascal}}]{McBride04}
\bibinfo{author}{\bibfnamefont{C.}~\bibnamefont{McBride}},
  \bibinfo{author}{\bibfnamefont{C.}~\bibnamefont{Vega}},
  \bibinfo{author}{\bibfnamefont{E.}~\bibnamefont{Sanz}}, \bibnamefont{and}
  \bibinfo{author}{\bibfnamefont{J.~L.~F.} \bibnamefont{Abascal}},
  \bibinfo{journal}{J. Chem. Phys.} \textbf{\bibinfo{volume}{121}},
  \bibinfo{pages}{11907} (\bibinfo{year}{2004}).

\bibitem[{\citenamefont{Marto\v{n}\'{a}k
  et~al.}(2004)\citenamefont{Marto\v{n}\'{a}k, Donadio, and
  Parrinello}}]{mdp_prl}
\bibinfo{author}{\bibfnamefont{R.}~\bibnamefont{Marto\v{n}\'{a}k}},
  \bibinfo{author}{\bibfnamefont{D.}~\bibnamefont{Donadio}}, \bibnamefont{and}
  \bibinfo{author}{\bibfnamefont{M.}~\bibnamefont{Parrinello}},
  \bibinfo{journal}{Phys. Rev. Lett.} \textbf{\bibinfo{volume}{92}},
  \bibinfo{pages}{225702} (\bibinfo{year}{2004}).

\bibitem[{\citenamefont{Lindahl et~al.}(2001)\citenamefont{Lindahl, Hess, and
  van~der Spoel}}]{gromacs}
\bibinfo{author}{\bibfnamefont{E.}~\bibnamefont{Lindahl}},
  \bibinfo{author}{\bibfnamefont{B.}~\bibnamefont{Hess}}, \bibnamefont{and}
  \bibinfo{author}{\bibfnamefont{D.}~\bibnamefont{van~der Spoel}},
  \bibinfo{journal}{J. Mol. Mod.} \textbf{\bibinfo{volume}{7}},
  \bibinfo{pages}{306} (\bibinfo{year}{2001}).

\bibitem[{\citenamefont{Parrinello and Rahman}(1980)}]{pr}
\bibinfo{author}{\bibfnamefont{M.}~\bibnamefont{Parrinello}} \bibnamefont{and}
  \bibinfo{author}{\bibfnamefont{A.}~\bibnamefont{Rahman}},
  \bibinfo{journal}{Phys. Rev. Lett.} \textbf{\bibinfo{volume}{45}},
  \bibinfo{pages}{1196} (\bibinfo{year}{1980}).

\bibitem[{\citenamefont{Berendsen et~al.}(1984)\citenamefont{Berendsen, Postma,
  van Gunsteren, DiNola, and Haak}}]{berendsen}
\bibinfo{author}{\bibfnamefont{H.~J.~C.} \bibnamefont{Berendsen}},
  \bibinfo{author}{\bibfnamefont{J.~P.~M.} \bibnamefont{Postma}},
  \bibinfo{author}{\bibfnamefont{W.~F.} \bibnamefont{van Gunsteren}},
  \bibinfo{author}{\bibfnamefont{A.}~\bibnamefont{DiNola}}, \bibnamefont{and}
  \bibinfo{author}{\bibfnamefont{J.~R.} \bibnamefont{Haak}},
  \bibinfo{journal}{J. Chem. Phys.} \textbf{\bibinfo{volume}{81}},
  \bibinfo{pages}{3684} (\bibinfo{year}{1984}).

\bibitem[{\citenamefont{Essmann et~al.}(1995)\citenamefont{Essmann, Perera,
  Berkowitz, Darden, Lee, and Pedersen}}]{pme}
\bibinfo{author}{\bibfnamefont{U.}~\bibnamefont{Essmann}},
  \bibinfo{author}{\bibfnamefont{L.}~\bibnamefont{Perera}},
  \bibinfo{author}{\bibfnamefont{M.~L.} \bibnamefont{Berkowitz}},
  \bibinfo{author}{\bibfnamefont{T.}~\bibnamefont{Darden}},
  \bibinfo{author}{\bibfnamefont{H.}~\bibnamefont{Lee}}, \bibnamefont{and}
  \bibinfo{author}{\bibfnamefont{L.~G.} \bibnamefont{Pedersen}},
  \bibinfo{journal}{J. Chem. Phys.} \textbf{\bibinfo{volume}{103}},
  \bibinfo{pages}{8577} (\bibinfo{year}{1995}).

\bibitem[{\citenamefont{Buch et~al.}(1998)\citenamefont{Buch, Sandler, and
  Sadlej}}]{buch}
\bibinfo{author}{\bibfnamefont{V.}~\bibnamefont{Buch}},
  \bibinfo{author}{\bibfnamefont{P.}~\bibnamefont{Sandler}}, \bibnamefont{and}
  \bibinfo{author}{\bibfnamefont{J.}~\bibnamefont{Sadlej}},
  \bibinfo{journal}{J. Phys. Chem. B} \textbf{\bibinfo{volume}{102}},
  \bibinfo{pages}{8641} (\bibinfo{year}{1998}).

\bibitem[{\citenamefont{Jorgensen et~al.}(1983)\citenamefont{Jorgensen,
  Chandrasekhar, Madura, Impey, and Klein}}]{tip4p}
\bibinfo{author}{\bibfnamefont{W.~L.} \bibnamefont{Jorgensen}},
  \bibinfo{author}{\bibfnamefont{J.}~\bibnamefont{Chandrasekhar}},
  \bibinfo{author}{\bibfnamefont{J.~D.} \bibnamefont{Madura}},
  \bibinfo{author}{\bibfnamefont{R.~W.} \bibnamefont{Impey}}, \bibnamefont{and}
  \bibinfo{author}{\bibfnamefont{M.~L.} \bibnamefont{Klein}},
  \bibinfo{journal}{J. Chem. Phys.} \textbf{\bibinfo{volume}{79}},
  \bibinfo{pages}{926} (\bibinfo{year}{1983}).

\bibitem[{\citenamefont{Sanz et~al.}(2004)\citenamefont{Sanz, Vega, Abascal,
  and MacDowell}}]{sanz04}
\bibinfo{author}{\bibfnamefont{E.}~\bibnamefont{Sanz}},
  \bibinfo{author}{\bibfnamefont{C.}~\bibnamefont{Vega}},
  \bibinfo{author}{\bibfnamefont{J.~L.~F.} \bibnamefont{Abascal}},
  \bibnamefont{and} \bibinfo{author}{\bibfnamefont{L.~G.}
  \bibnamefont{MacDowell}}, \bibinfo{journal}{Phys. Rev. Lett.}
  \textbf{\bibinfo{volume}{92}}, \bibinfo{pages}{255701}
  (\bibinfo{year}{2004}).

\bibitem[{\citenamefont{Tanaka}(2000)}]{Tanaka2000}
\bibinfo{author}{\bibfnamefont{H.}~\bibnamefont{Tanaka}}, \bibinfo{journal}{J.
  Chem. Phys.} \textbf{\bibinfo{volume}{113}}, \bibinfo{pages}{11202}
  (\bibinfo{year}{2000}).

\bibitem[{\citenamefont{Klotz et~al.}(2003)\citenamefont{Klotz, Hamel, Loveday,
  Nelmes, and Guthrie}}]{klotz_recryst}
\bibinfo{author}{\bibfnamefont{S.}~\bibnamefont{Klotz}},
  \bibinfo{author}{\bibfnamefont{G.}~\bibnamefont{Hamel}},
  \bibinfo{author}{\bibfnamefont{J.}~\bibnamefont{Loveday}},
  \bibinfo{author}{\bibfnamefont{R.}~\bibnamefont{Nelmes}}, \bibnamefont{and}
  \bibinfo{author}{\bibfnamefont{M.}~\bibnamefont{Guthrie}},
  \bibinfo{journal}{Z. Kristallogr.} \textbf{\bibinfo{volume}{218}},
  \bibinfo{pages}{117} (\bibinfo{year}{2003}).

\bibitem[{\citenamefont{Mishima and Stanley}(1998)}]{entropy}
\bibinfo{author}{\bibfnamefont{O.}~\bibnamefont{Mishima}} \bibnamefont{and}
  \bibinfo{author}{\bibfnamefont{H.~E.} \bibnamefont{Stanley}},
  \bibinfo{journal}{Nature (London)} \textbf{\bibinfo{volume}{396}},
  \bibinfo{pages}{329} (\bibinfo{year}{1998}).

\bibitem[{\citenamefont{Trachenko and Dove}(2003{\natexlab{a}})}]{trachenko1}
\bibinfo{author}{\bibfnamefont{K.}~\bibnamefont{Trachenko}} \bibnamefont{and}
  \bibinfo{author}{\bibfnamefont{M.~T.} \bibnamefont{Dove}},
  \bibinfo{journal}{Phys. Rev. B} \textbf{\bibinfo{volume}{67}},
  \bibinfo{pages}{064107} (\bibinfo{year}{2003}{\natexlab{a}}).

\bibitem[{\citenamefont{D\'{a}vila et~al.}(2003)\citenamefont{D\'{a}vila,
  Caturla, Kubota, Sadigh, de~la Rubia, Shackelford, Risbud, and
  Garofalini}}]{davila}
\bibinfo{author}{\bibfnamefont{L.~P.} \bibnamefont{D\'{a}vila}},
  \bibinfo{author}{\bibfnamefont{M.-J.} \bibnamefont{Caturla}},
  \bibinfo{author}{\bibfnamefont{A.}~\bibnamefont{Kubota}},
  \bibinfo{author}{\bibfnamefont{B.}~\bibnamefont{Sadigh}},
  \bibinfo{author}{\bibfnamefont{T.~D.} \bibnamefont{de~la Rubia}},
  \bibinfo{author}{\bibfnamefont{J.~F.} \bibnamefont{Shackelford}},
  \bibinfo{author}{\bibfnamefont{S.~H.} \bibnamefont{Risbud}},
  \bibnamefont{and} \bibinfo{author}{\bibfnamefont{S.~H.}
  \bibnamefont{Garofalini}}, \bibinfo{journal}{Phys. Rev. Lett.}
  \textbf{\bibinfo{volume}{91}}, \bibinfo{pages}{205501}
  (\bibinfo{year}{2003}).

\bibitem[{\citenamefont{Yuan and Cormack}(2002)}]{yuan02}
\bibinfo{author}{\bibfnamefont{X.~L.} \bibnamefont{Yuan}} \bibnamefont{and}
  \bibinfo{author}{\bibfnamefont{A.~N.} \bibnamefont{Cormack}},
  \bibinfo{journal}{Comp. Mat. Sci.} \textbf{\bibinfo{volume}{24}},
  \bibinfo{pages}{343} (\bibinfo{year}{2002}).

\bibitem[{\citenamefont{Raiteri et~al.}(2004)\citenamefont{Raiteri, Laio, and
  Parrinello}}]{paolo04}
\bibinfo{author}{\bibfnamefont{P.}~\bibnamefont{Raiteri}},
  \bibinfo{author}{\bibfnamefont{A.}~\bibnamefont{Laio}}, \bibnamefont{and}
  \bibinfo{author}{\bibfnamefont{M.}~\bibnamefont{Parrinello}},
  \bibinfo{journal}{Phys. Rev. Lett.} \textbf{\bibinfo{volume}{93}},
  \bibinfo{pages}{87801} (\bibinfo{year}{2004}).

\bibitem[{\citenamefont{Finney et~al.}(2002{\natexlab{b}})\citenamefont{Finney,
  Hallbrucker, Kohl, Soper, and Bowron}}]{finney_hlda}
\bibinfo{author}{\bibfnamefont{J.~L.} \bibnamefont{Finney}},
  \bibinfo{author}{\bibfnamefont{A.}~\bibnamefont{Hallbrucker}},
  \bibinfo{author}{\bibfnamefont{I.}~\bibnamefont{Kohl}},
  \bibinfo{author}{\bibfnamefont{A.~K.} \bibnamefont{Soper}}, \bibnamefont{and}
  \bibinfo{author}{\bibfnamefont{D.~T.} \bibnamefont{Bowron}},
  \bibinfo{journal}{Phys. Rev. Lett.} \textbf{\bibinfo{volume}{88}},
  \bibinfo{pages}{225503} (\bibinfo{year}{2002}{\natexlab{b}}).

\bibitem[{\citenamefont{Soper}(1996)}]{epsr}
\bibinfo{author}{\bibfnamefont{A.~K.} \bibnamefont{Soper}},
  \bibinfo{journal}{Chem. Phys.} \textbf{\bibinfo{volume}{202}},
  \bibinfo{pages}{295} (\bibinfo{year}{1996}).

\bibitem[{\citenamefont{Saitta et~al.}(2004)\citenamefont{Saitta, Str\"assle,
  Rousse, Hamel, Klotz, Nelmes, and Loveday}}]{saitta}
\bibinfo{author}{\bibfnamefont{A.~M.} \bibnamefont{Saitta}},
  \bibinfo{author}{\bibfnamefont{T.}~\bibnamefont{Str\"assle}},
  \bibinfo{author}{\bibfnamefont{G.}~\bibnamefont{Rousse}},
  \bibinfo{author}{\bibfnamefont{G.}~\bibnamefont{Hamel}},
  \bibinfo{author}{\bibfnamefont{S.}~\bibnamefont{Klotz}},
  \bibinfo{author}{\bibfnamefont{R.~J.} \bibnamefont{Nelmes}},
  \bibnamefont{and} \bibinfo{author}{\bibfnamefont{J.~S.}
  \bibnamefont{Loveday}}, \bibinfo{journal}{J. Chem. Phys.}
  \textbf{\bibinfo{volume}{121}}, \bibinfo{pages}{8430} (\bibinfo{year}{2004}).

\bibitem[{\citenamefont{Wilding}(1997)}]{finite-size-scaling}
\bibinfo{author}{\bibfnamefont{N.~B.} \bibnamefont{Wilding}},
  \bibinfo{journal}{J. Phys.: Condens. Matter} \textbf{\bibinfo{volume}{9}},
  \bibinfo{pages}{585} (\bibinfo{year}{1997}).

\bibitem[{\citenamefont{van Duijneveldt and Frenkel}(1992)}]{frenkel}
\bibinfo{author}{\bibfnamefont{J.~S.} \bibnamefont{van Duijneveldt}}
  \bibnamefont{and} \bibinfo{author}{\bibfnamefont{D.}~\bibnamefont{Frenkel}},
  \bibinfo{journal}{J. Chem. Phys.} \textbf{\bibinfo{volume}{96}},
  \bibinfo{pages}{4655} (\bibinfo{year}{1992}).

\bibitem[{\citenamefont{Zachariasen}(1932)}]{zach_cnr32}
\bibinfo{author}{\bibfnamefont{W.~H.} \bibnamefont{Zachariasen}},
  \bibinfo{journal}{J. Amer. Chem. Soc.} \textbf{\bibinfo{volume}{54}},
  \bibinfo{pages}{3841} (\bibinfo{year}{1932}).

\bibitem[{\citenamefont{Marians and Hobbs}(1990{\natexlab{a}})}]{hobbs90}
\bibinfo{author}{\bibfnamefont{C.~S.} \bibnamefont{Marians}} \bibnamefont{and}
  \bibinfo{author}{\bibfnamefont{L.~W.} \bibnamefont{Hobbs}},
  \bibinfo{journal}{J. non-Cryst. Solids} \textbf{\bibinfo{volume}{119}},
  \bibinfo{pages}{269} (\bibinfo{year}{1990}{\natexlab{a}}).

\bibitem[{\citenamefont{Grimsditch}(1984)}]{grimsditch}
\bibinfo{author}{\bibfnamefont{M.}~\bibnamefont{Grimsditch}},
  \bibinfo{journal}{Phys. Rev. Lett.} \textbf{\bibinfo{volume}{52}},
  \bibinfo{pages}{2379} (\bibinfo{year}{1984}).

\bibitem[{\citenamefont{Hemley et~al.}(1986)\citenamefont{Hemley, Mao, Bell,
  and Mysen}}]{hemley86}
\bibinfo{author}{\bibfnamefont{R.~J.} \bibnamefont{Hemley}},
  \bibinfo{author}{\bibfnamefont{H.~K.} \bibnamefont{Mao}},
  \bibinfo{author}{\bibfnamefont{P.~M.} \bibnamefont{Bell}}, \bibnamefont{and}
  \bibinfo{author}{\bibfnamefont{B.~O.} \bibnamefont{Mysen}},
  \bibinfo{journal}{Phys. Rev. Lett.} \textbf{\bibinfo{volume}{57}},
  \bibinfo{pages}{747} (\bibinfo{year}{1986}).

\bibitem[{\citenamefont{Williams and Jeanloz}(1988)}]{williams88}
\bibinfo{author}{\bibfnamefont{Q.}~\bibnamefont{Williams}} \bibnamefont{and}
  \bibinfo{author}{\bibfnamefont{R.}~\bibnamefont{Jeanloz}},
  \bibinfo{journal}{Science} \textbf{\bibinfo{volume}{239}},
  \bibinfo{pages}{902} (\bibinfo{year}{1988}).

\bibitem[{\citenamefont{Lacks}(2000)}]{lacks00}
\bibinfo{author}{\bibfnamefont{D.}~\bibnamefont{Lacks}},
  \bibinfo{journal}{Phys. Rev. Lett.} \textbf{\bibinfo{volume}{84}},
  \bibinfo{pages}{4629} (\bibinfo{year}{2000}).

\bibitem[{\citenamefont{Trachenko and Dove}(2003{\natexlab{b}})}]{trachenko2}
\bibinfo{author}{\bibfnamefont{K.}~\bibnamefont{Trachenko}} \bibnamefont{and}
  \bibinfo{author}{\bibfnamefont{M.~T.} \bibnamefont{Dove}},
  \bibinfo{journal}{Phys. Rev. B} \textbf{\bibinfo{volume}{67}},
  \bibinfo{pages}{212203} (\bibinfo{year}{2003}{\natexlab{b}}).

\bibitem[{\citenamefont{Huang and Kieffer}(2004{\natexlab{a}})}]{huang1}
\bibinfo{author}{\bibfnamefont{L.}~\bibnamefont{Huang}} \bibnamefont{and}
  \bibinfo{author}{\bibfnamefont{J.}~\bibnamefont{Kieffer}},
  \bibinfo{journal}{Phys. Rev. B} \textbf{\bibinfo{volume}{69}},
  \bibinfo{pages}{224203} (\bibinfo{year}{2004}{\natexlab{a}}).

\bibitem[{\citenamefont{Huang and Kieffer}(2004{\natexlab{b}})}]{huang2}
\bibinfo{author}{\bibfnamefont{L.}~\bibnamefont{Huang}} \bibnamefont{and}
  \bibinfo{author}{\bibfnamefont{J.}~\bibnamefont{Kieffer}},
  \bibinfo{journal}{Phys. Rev. B} \textbf{\bibinfo{volume}{69}},
  \bibinfo{pages}{224204} (\bibinfo{year}{2004}{\natexlab{b}}).

\bibitem[{\citenamefont{Stixrude and Bukowinski}(1991)}]{stixrude_amorph}
\bibinfo{author}{\bibfnamefont{L.}~\bibnamefont{Stixrude}} \bibnamefont{and}
  \bibinfo{author}{\bibfnamefont{M.~S.~T.} \bibnamefont{Bukowinski}},
  \bibinfo{journal}{Phys. Rev. B} \textbf{\bibinfo{volume}{44}},
  \bibinfo{pages}{2523} (\bibinfo{year}{1991}).

\bibitem[{\citenamefont{Grimsditch et~al.}(1996)\citenamefont{Grimsditch,
  Polian, and Wright}}]{grimsditch_b2o3}
\bibinfo{author}{\bibfnamefont{M.}~\bibnamefont{Grimsditch}},
  \bibinfo{author}{\bibfnamefont{A.}~\bibnamefont{Polian}}, \bibnamefont{and}
  \bibinfo{author}{\bibfnamefont{A.~C.} \bibnamefont{Wright}},
  \bibinfo{journal}{Phys. Rev. B} \textbf{\bibinfo{volume}{54}},
  \bibinfo{pages}{152} (\bibinfo{year}{1996}).

\bibitem[{\citenamefont{Marians and Hobbs}(1990{\natexlab{b}})}]{marians_cryst}
\bibinfo{author}{\bibfnamefont{C.~S.} \bibnamefont{Marians}} \bibnamefont{and}
  \bibinfo{author}{\bibfnamefont{L.~W.} \bibnamefont{Hobbs}},
  \bibinfo{journal}{J. Non-Cryst. Solids} \textbf{\bibinfo{volume}{124}},
  \bibinfo{pages}{242} (\bibinfo{year}{1990}{\natexlab{b}}).

\end{thebibliography}

\end{document}